\documentclass[a4paper,11pt]{article}

\usepackage{amsmath}
\usepackage{amssymb}
\usepackage{graphicx}
\usepackage{bbm}
\usepackage{cite}

\newcommand{\be}{\begin{equation}}  
\newcommand{\ee}{\end{equation}}  
\newcommand{\ol}[1]{\overline{#1}}
\newcommand{\hc}{+\,\mathrm{h.c.}}

\newcommand{\SU}[1]{\ensuremath{\mathrm{SU}(#1)}}

\newcommand{\tr}{\operatorname{tr}}

\newcommand{\delGM}{\ensuremath{\Delta_{\widehat m_Z^2}^{\text{\tiny GM}}}}
\newcommand{\mgm}{\ensuremath{m_{\text{\tiny GM}}}}
\newcommand{\mS}{\ensuremath{m_{\text{\tiny S}}}}

\title{
\vspace{-4.5ex}
{\normalsize \flushright
DESY 12--009\\
January 2012\\[10ex]
}
\textbf{The Fermi scale as a focus point of high-scale gauge mediation}
\vspace{2ex}
}

\author{F.~Br\"ummer and W.~Buchm\"uller\\[1ex]
\textit{\normalsize Deutsches Elektronen-Synchrotron DESY,}\\
\textit{\normalsize Notkestra\ss e 85, D-22607 Hamburg, Germany}
\vspace{3ex}
}

\date{}

\begin{document}

\maketitle

\begin{abstract}
\noindent 
We consider the minimal supersymmetric Standard Model with large
scalar and gaugino mass terms at the GUT scale, which are generated
predominantly by gauge-mediated supersymmetry breaking. For certain ratios 
of GUT-scale masses, determined by the messenger indices, large 
radiative corrections lead to a small electroweak scale 
in a way which resembles the well-known focus point mechanism. The
Fermi scale, the gravitino mass and the higgsino masses are of comparable
size. For a Higgs mass of about 124 GeV all other superparticles have
masses outside the reach of the LHC.

\end{abstract}

\section{Introduction}

The minimal supersymmetric Standard Model (MSSM) is increasingly coming under
pressure from the results of LHC searches. The non-observation of squarks and
gluinos has defied expectations to find them just above the pre-LHC exclusion
bounds. In most generic scenarios they must now be heavier than about a TeV.
Perhaps more severely, evidence for a Higgs boson from both ATLAS \cite{atlas}
and CMS \cite{cms} points to a Higgs mass of around $124-126$ GeV. Should this
evidence solidify, this would pose a serious naturalness problem for the MSSM,
or at least for many top-down scenarios which assume a ``great desert'' between
the TeV scale and the GUT scale. The reason is that such a large Higgs mass must
be due to large loop corrections involving multi-TeV third-generation squarks,
or near-maximal squark mixing. On the other hand, the third-generation squark
mass and gaugino mass contributions dominate the renormalization group (RG) 
evolution of the Higgs soft masses. The natural size of the parameters governing 
the Higgs potential is therefore of the order of these large soft masses. If the
electroweak symmetry breaking (EWSB) scale is to come out smaller than the soft
mass scale by an order of magnitude or more, this requires large cancellations,
and a corresponding fine-tuning of the GUT-scale boundary conditions.

It is conceivable that a ``little hierarchy'' of this kind is a consequence of
certain relations among the GUT-scale soft terms, or that such relations at 
least reduce the degree of fine-tuning. The best-known example is perhaps 
focus point supersymmetry \cite{Chan:1997bi,Feng:1999mn, Feng:2011aa,Akula:2011jx},
which rests on the observation that, for comparatively small gaugino masses and at least
moderately large $\tan\beta$, the scalar soft mass contributions to the EWSB
order parameter $m_Z^2$ cancel through RG running. For this to happen, 
all scalar soft masses should be given by some universal $m_0$ at the GUT scale. 
The cancellation is then quite insensitive to the actual value of $m_0$ (although, of 
course, it only holds at a finite level of precision). Therefore, it has been argued 
that increasing $m_0$ to several TeV does not render the model any more unnatural. 
As the main source of fine-tuning is often the gluino mass rather than the scalar
soft masses, there have also been searches for similarly favourable relations
between gaugino masses in models without gaugino mass unification
\cite{Abe:2007kf,Horton:2009ed} (see also \cite{Younkin:2012xx} for a related 
recent study).

This note serves to point out that focus point-like models may be constructed in
the framework of high-scale gauge mediation with split messenger multiplets, as
proposed in \cite{Brummer:2011yd}. In these models, soft gaugino and scalar
masses receive large gauge-mediated contributions. Gauge-mediated soft terms are
dictated by discrete parameters, namely the messenger indices. For certain models
 the messenger indices are such that the contributions to
$m_Z^2$ cancel between the various soft terms during RG running. The soft terms
also receive subdominant contributions from gravity mediation. Since the large
gauge-mediated contributions cancel, the EWSB scale is naturally set by the
subdominant gravity-mediated terms, and these can well be of the order of $m_Z^2$
without conflicting experiment. Our model thus shows that it is still possible
to reconcile the MSSM with the latest LHC results within a well-motivated GUT
framework. This can even be achieved without relying on large stop mixing,
which is the mechanism underlying most bottom-up attempts to maximize the
lightest Higgs mass in the MSSM, in light of the recent LHC results
\cite{Heinemeyer:2011aa, Arbey:2011ab, Draper:2011aa, Carena:2011aa,Akula:2011aa}.

\section{Framework}
Our models are inspired by GUT-scale string compactifications, which often
predict a large number of vector-like states in incomplete GUT multiplets. These
exotics should decouple close to the GUT scale in order to preserve
(approximate) gauge coupling unification. If they couple to the hidden sector,
they will serve as messenger fields for gauge-mediated supersymmetry breaking.
Messenger loops generate large soft masses for the gauginos and scalars. Other 
soft terms and the higgsino mass $\mu$ will be induced by
Planck-suppressed operators; they will be smaller but still relevant.

For definiteness, consider a model with $N_3$ copies
$(\Phi_{a_3}^{(3)},\widetilde\Phi_{a_3}^{(3)})$ of $({\bf 3},{\bf
1})_0\oplus(\ol{\bf 3},{\bf 1})_0$, $N_2$ copies
$(\Phi_{a_2}^{(2)},\widetilde\Phi_{a_2}^{(2)})$ of $({\bf 1},{\bf
2})_0\oplus({\bf 1},{\bf 2})_0$, and $N_1$ copies
$(\Phi_{a_1}^{(1)},\widetilde\Phi_{a_1}^{(1)})$ of $({\bf 1},{\bf
1})_1\oplus({\bf 1},{\bf 1})_{-1}$ as messenger supermultiplets.\footnote{Compare for
example the model in Section 3.1 of \cite{Brummer:2011yd}, which effectively has
 $N_1=10$, $N_2=16$, and $N_3=6$.} 
Suppose that
they couple with equal strength to a goldstino background field $X=M+\theta^2F$,
such that the superpotential is
\be\label{eq:messengerW}
W=\sum_{i=1}^3\sum_{a_i=1}^{N_i} X\Phi_{a_i}^{(i)}\widetilde\Phi_{a_i}^{(i)}\,.
\ee
For $M\lesssim M_{\rm GUT}$, we define 
\be
\mgm\equiv\frac{g^2}{16\pi^2}\frac{F}{M}
\ee
to be the typical gauge-mediated contribution to the soft masses per
messenger pair, where $g$ is the unified gauge coupling.  
The one-loop gauge-mediated gaugino masses at the scale $M$ are
\be\begin{split}\label{eq:gauginos}
M_1&=\frac{6}{5}\,N_1\,\mgm\,,\\
M_2&=N_2\,\mgm\,,\\
M_3&=N_3\,\mgm\,.
\end{split}
\ee
The two-loop scalar soft masses are
\be\begin{split}\label{eq:scalars}
m_Q^2&=\left(\frac{8}{3}N_3+\frac{3}{2}
N_2+\frac{1}{25}N_1\right)\mgm^2,\\
m_U^2&=\left(\frac{8}{3}N_3+\frac{16}{25}
N_1\right)\mgm^2\,,\\
m_D^2&=\left(\frac{8}{3}N_3+\frac{4}{25}
N_1\right)\mgm^2\,,\\
m_E^2&=\left(\frac{36}{25}
N_1\right)\mgm^2\,,\\
m_{H_{u,d}}^2=m_L^2&=\left(\frac{3}{2}
N_2+\frac{9}{25}N_1\right)\mgm^2\,.
\end{split}
\ee
Here we have assumed that the running between $M_{\rm GUT}$ and $M$ can be
neglected. For $\mgm$ to be ${\cal O}(10^{2-3})$ GeV, we need 
$F\simeq (\text{few}\times
10^{10}\,\text{GeV})^2$. This implies that gravity mediation is not negligible,
\be
\frac{F}{\sqrt{3} M_{\rm Pl}}=m_{3/2}\simeq \mgm\,.
\ee 
However, gaugino and scalar soft masses will be dominated by the gauge-mediated
contributions if the $N_i$ are moderately large.
The $A$, $B_\mu$ and $\mu$ parameters, which are small or absent in minimal
gauge mediation, are induced by gravity mediation and will be of the order
$m_{3/2}$. Unlike in low-scale gauge mediation models, the flavour problem is
not automatically solved; we therefore require the gravity-mediated soft terms
to be approximately flavour preserving. Incidentally, the particle spectrum in
our models will be rather heavy, which of course also helps with evading
flavour constraints.

\section{A focus point from high-scale gauge mediation}
Recall that at large $\tan\beta$, the $Z$ mass
is given by\footnote{Here and in the following, we use the symbol 
$\widehat m_Z$ for the quantity calculated from the MSSM Higgs potential, reserving 
$m_Z$ for the actual physical pole mass $m_Z=91.2$ GeV.}
\be
-\frac{\widehat m_Z^2}{2}=\left.\left(|\mu|^2+m_{H_u}^2\right)\right|_{\mS}
\ee
where $\mu$ and $m_{H_u}$ are the running masses at the scale $\mS$ where the
Higgs potential is minimized (as usual taken to be $\mS=\sqrt{m_{\tilde
t_1}m_{\tilde t_2}}$). Unlike the Higgsino mass $\mu$, the Higgs soft mass
$m_{H_u}^2$ receives large radiative corrections through its RG evolution. If
the gaugino and scalar soft terms are ${\cal O}(\text{few TeV})$ while $\mu$ is
${\cal O}(100\,{\rm GeV})$, then the UV-scale value of $m_{H_u}^2$ must
approximately cancel against these radiative corrections for $m_{H_u}^2$ to end
up small (and negative) at $\mS$. We can estimate $\widehat m_Z$ in terms of the UV-scale
parameters as
\be\begin{split}\label{eq:numericMZ}
\widehat m_Z^2=\,&\Bigl(2.25\,M_3^2-0.45\,M_2^2-0.01\,M_1^2+0.19\,M_2 M_3+0.03\,M_1 M_3\\
&+0.74\,m_{U}^2+0.65\,m_Q^2-0.04\,m_D^2-1.32\,m_{H_u}^2-0.09\,m_{H_d}^2\\
&+0.19\,A_0^2-0.40\,A_0 M_3-0.11\,A_0 M_2-0.02\,A_0 M_1\\
&-1.42\,|\mu|^2\left.\Bigr)\right|_{M}\,.
\end{split}\ee
Here the functional form of the RHS follows from the form of the RG equations
and from dimensional analysis.
 The coefficients have been determined assuming a universal trilinear parameter
$A_0$, and with $M=10^{16}$ GeV, $\mS=3.5$ TeV, $g^2(M)=0.484$, $y_t(M)=0.55$,
$y_b(M)=0.4$, and $y_\tau(M)=0.5$; these are typical values for a heavy spectrum
at $\tan\beta\approx 50$. We have used two-loop RG equations for the gauge and
Yukawa couplings and for the gaugino masses, and one-loop RG equations
otherwise. Terms with coefficients smaller than $0.01$ have been omitted.

From Eq.~\eqref{eq:numericMZ} one can immediately read off that, as is well
known, the electroweak scale is most sensitive to the gluino mass among all soft
parameters. The second line also illustrates the original focus point
scenario \cite{Chan:1997bi,Feng:1999mn}: If $M_{1,2,3}$ and $\mu$ are small, and if there 
is a universal GUT-scale scalar soft mass $m_0$, then $m_0$ can be rather large 
without excessive fine-tuning. Pictorially speaking, the RG trajectories of 
$m_{H_u}^2$ for various choices of $m_0$ ``focus'' to always cross zero close to 
the Fermi scale. 

In our class of models, the gaugino masses are by no means small. Any focus
point-like cancellation will have to involve all of the dominant gauge-mediated
terms, at least including $M_3$, $M_2$, $m_{H_u}^2$, $m_U^2$ and $m_Q^2$.
Neglecting gravity-mediated contributions for the moment, in particular $\mu$ and 
$A_0$, we can insert Eqns.~\eqref{eq:gauginos} and
\eqref{eq:scalars} into Eq.~\eqref{eq:numericMZ} to find the contribution to
$\widehat m_Z^2$ which is purely due to the gauge-mediated soft terms:
\be\begin{split}\label{eq:numericNs}
\delGM=\,&\bigl(2.25\,N_3^2-0.45\,N_2^2-0.01\,N_1^2\\
&+0.19\,N_2N_3+0.04\,N_1N_3\\
&+3.80\,N_3 -1.16\,N_2-0.01\,N_1\,\bigr)\,\mgm^2\,.
\end{split}
\ee
A small contribution to $\widehat m_Z^2$ results for certain favourable ratios
of $N_2$ and $N_3$, while the influence of $N_1$ is subdominant. 
Indeed there is not even a reason for $N_1$ to be
integral, since a generic model may contain several hypercharged messengers with
various rational hypercharges. Thus, setting $N_1=N_2$ for simplicity, we find that
$\delGM$ is exceptionally small for
\be\label{eq:focus1}
N_2=23,\qquad N_3=9
\ee
or
\be\label{eq:focus2}
N_2=28,\qquad N_3=11\,.
\ee
More precisely, for $N_1=N_2=23$ and $N_3=9$ we get a contribution to $\widehat m_Z^2$
which is of the order of $\mgm$, while the actual soft masses are
larger by an order of magnitude:
\be\label{eq:mZgauge1}
\delGM= 0.6\,\mgm^2\,.
\ee 
Similarly, for $N_1=N_2=28$ and $N_3=11$ one obtains
\be\label{eq:mZgauge2}
\delGM= 1.3\,\mgm^2\,.
\ee 

\begin{figure}
 \begin{center}
\includegraphics[width=.67\textwidth]{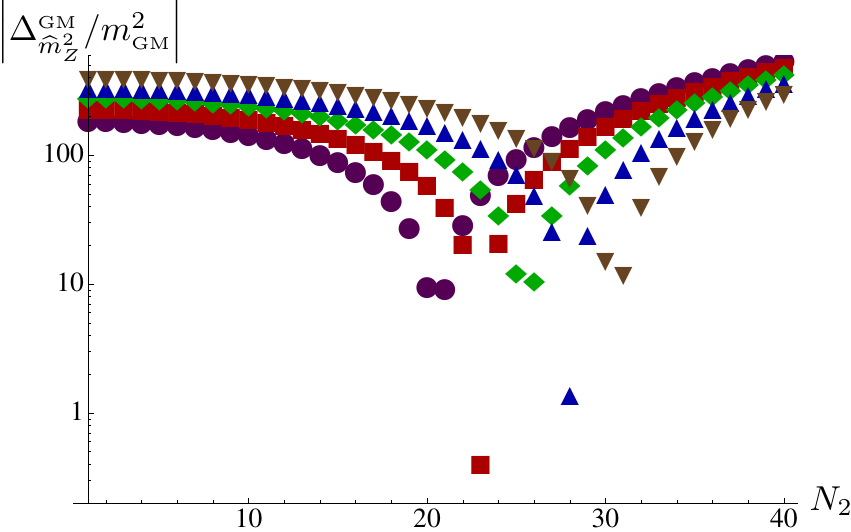}
\end{center}
\caption{The ratio $\delGM/\mgm^2$ for varying $N_2$, with $N_1=N_2$ and $N_3=8$ 
(purple circles), $N_3=9$ (red squares), $N_3=10$ (green diamonds), $N_3=11$ 
(blue triangles) and $N_3=12$ (brown inverted triangles). On the left branch of
each curve, electroweak symmetry is broken, while it is unbroken on the right
branch. Note the exceptionally small ratios at
$(N_2,N_3)=(23,9)$ and $(28,11)$.
}\label{fig:cancellation}
\end{figure}

The cancellation in the gauge-mediated contribution to ${\widehat m_Z}^2$ is 
illustrated in Fig.~\ref{fig:cancellation}, where we have logarithmically 
plotted $\delGM$
in units of $\mgm^2$ for various choices of $N_3$ and $N_2$, after
setting $N_1=N_2$. For positive ${\widehat m_Z}^2$ (towards smaller $N_2$) 
the electroweak symmetry
is broken and the magnitude of ${\widehat m_Z}^2$ represents the electroweak
scale. If ${\widehat m_Z}^2$ is negative (towards larger $N_2$), the electroweak 
symmetry is unbroken and its magnitude corresponds to the running Higgs mass 
squared at the TeV scale. The plot illustrates the well-known
problem of quadratic divergences \cite{Veltman:1980mj}: Generically, the
scale of symmetry breaking is given by the cutoff of the theory, in our case
the scale of grand unification. In supersymmetric theories, the cutoff is
replaced by the supersymmetry breaking mass terms, which reduces the
``hierarchy problem'' to the ``little hierarchy problem''. Hence the Higgs mass
squared at a TeV is typically of the order $N_{2,3}^2 \mgm^2$. However,
for certain relations between the superparticle mass terms, corresponding to 
particular messenger indices, cancellations occur and the scale of 
electroweak symmetry breaking can be of the order $\mgm^2$. 

In general the soft terms will also receive contributions from gravity
mediation. This is especially problematic for $M_3$, since changing the gaugino
mass ratios resulting from Eq.~\eqref{eq:focus1} or Eq.~\eqref{eq:focus2}
by more than a few percent will immediately spoil the cancellation between the 
soft terms in Eq.~\eqref{eq:numericMZ}, in view of the large coefficient of $M_3^2$. 
The dangerous leading gravity-mediated contribution to gaugino masses comes from 
the operator
\be\label{eq:gravgauginomass}
{\cal L}\supset\int d^2\theta\frac{X}{M_{\rm Pl}}\tr W^\alpha W_\alpha\,\hc
\ee
Likewise, the $A_0 M_i$ terms in Eq.~\eqref{eq:numericMZ} are $N_i$-enhanced and
therefore potentially dangerous. The leading operators inducing $A$-terms are
\be\label{eq:tril}\begin{split}
{\cal L}\supset&\int d^2\theta\frac{X}{M_{\rm Pl}}\left(H_u QU+H_d QD+H_d LE\right)\hc\\
&+\int d^4\theta\frac{X}{M_{\rm Pl}}\left(|H_u|^2+|H_d|^2+|Q|^2+|U|^2+|D|^2+|L|^2+|E|^2\right)\hc\\
\end{split}
\ee

All operators of Eqns.~\eqref{eq:gravgauginomass} and \eqref{eq:tril} can be 
forbidden by a symmetry under which $X$ is charged (and which is spontaneously 
broken by the VEV of $X$). By forbidding gravity-mediated $A$-terms, we are also
removing a major source of dangerous flavour-changing neutral currents. 
The Giudice-Masiero term
\be
{\cal L}\supset\int d^4\theta\frac{X^\dag}{M_{\rm Pl}}H_u H_d\,\hc\,
\ee
is forbidden as well, but an effective $\mu$ parameter of the order $m_{3/2}$ can
still arise from superpotential couplings after SUSY breaking.
More precisely, in a SUSY-breaking Minkowski vacuum, the VEV of the superpotential 
is of the order $W_0\simeq F\,M_{\rm Pl}$. A continuous or discrete $R$-symmetry 
forbids a bare $\mu$ term while allowing for an operator
\be 
{\cal L}\supset\int d^2\theta \frac{W_0}{M_{\rm Pl}^2} H_u H_d\hc\,,
\ee
which generates an effective $\mu\simeq F/M_{\rm Pl}$ \cite{Casas:1992mk, Buchmuller:2008uq, Brummer:2010fr}.  
Note that the operator
\be
{\cal L}\supset\int d^4\theta\frac{X^\dag X}{M_{\rm Pl}^2}H_u H_d\hc
\ee
which induces $B_\mu$ is allowed in any case, as are the operators for 
gravity-mediated contributions to scalar soft masses
\be
{\cal L}\supset \int d^4\theta\frac{X^\dag X}{M_{\rm Pl}^2}\left(
|H_u|^2+|H_d|^2+|Q|^2+|U|^2+|D|^2+|L|^2+|E|^2\right)\,.
\ee
The latter will give small corrections to the soft masses of Eqns.~\eqref{eq:scalars},
without much affecting the focus point cancellation.

The gaugino masses also receive corrections at two-loop and higher orders.
However, it was shown in \cite{ArkaniHamed:1998kj} that for large messenger
masses and multiplicities, these corrections tend to be very small.

Without radiative corrections, the lightest Higgs mass would be bounded from
above by $m_Z$, with the bound saturated at large $\tan\beta$. Including the
one-loop corrections from the top-stop sector, at large $\tan\beta$ we have
(see e.g.~\cite{Martin:1997ns})
\be\label{eq:mZmh0}
\begin{split}
\frac{m_{Z}^2}{m_{h_0}^2}=\Biggl[&1+\frac{3}{2\pi^2}\frac{y_t^4}{g_1^2+g_2^2}
\Biggl(\log\frac{\mS^2}{m_t^2}
+\frac{A_t^2}{\mS^2}\left(1-\frac{A_t^2}{12\,\mS^2}
\right)\Biggr)\Biggr]^{-1}\,,
\end{split}
\ee
where as before $\mS^2=m_{\tilde t_1} m_{\tilde t_2}$. Note that a Higgs mass
around 125 GeV requires radiative corrections of the same size as the 
tree-level mass.

\begin{figure}
 \begin{center}
\includegraphics[width=.67\textwidth]{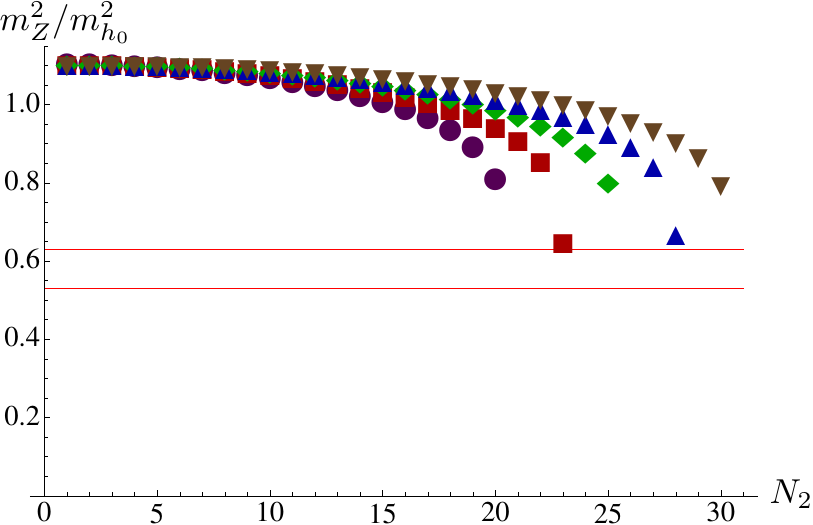}
\end{center}
\caption{The ratio $m_Z^2/m_{h_0}^2$, estimated from Eq.~\eqref{eq:mZmh0}
with gauge-mediated contributions only, for various $N_2$ and $N_3$ as in 
Fig.~\ref{fig:cancellation}. The red lines 
correspond to  $m_{h_0}=115$ GeV (upper line) and $m_{h_0}=125$ GeV (lower line).
While the actual Higgs masses (as computed by \texttt{SOFTSUSY} or 
\texttt{FeynHiggs}) tend to be even a few GeV higher, such that the $124-126$ GeV
region can be reached, the figure still shows that for the messenger indices
of Eqns.~\eqref{eq:focus1} and \eqref{eq:focus2}, the 
enhancement of $m_{h_0}$ above $m_Z$ is particularly strong.}\label{fig:mh0mZ}
\end{figure}

Exceptionally large radiative corrections to the Higgs mass point to special
properties of the Higgs-top-stop system, in our case stop masses much larger
than the scale of electroweak symmetry breaking. As discussed above, this
situation can be realized if the gauge-mediated contributions to squark masses
are large and the electroweak scale emerges as a focus point. 
As we did for $\widehat m_Z^2$ in Eq.~\eqref{eq:numericMZ},
we can now again express all dimensionful parameters on the RHS of 
Eq.~\eqref{eq:mZmh0} as functions of the GUT-scale soft terms, with the 
coefficients determined from the RG equations (for instance, setting 
$m_t^2=\frac{2 y_t^2}{g_1^2+g_2^2} \widehat m_Z^2$ with 
${\widehat m_Z}^2$ taken from Eq.~\eqref{eq:numericMZ}). The resulting ratio 
$m_Z^2/m_{h_0}^2$ is plotted
in Fig.~\ref{fig:mh0mZ} for various messenger numbers, again using only the
gauge-mediated contributions as given by Eqns.~\eqref{eq:gauginos} and 
\eqref{eq:scalars}. For generic messenger indices, $m^2_{h_0}$ exceeds
$m^2_Z$  by less than 20\%. On the contrary, for messenger indices 
yielding the Fermi scale as a focus point, $m^2_{h_0}$ exceeds
$m^2_Z$  by more than 70\%.

\section{Phenomenology}

Let us first consider a model where the messenger numbers are given by
$N_1=N_2=23$, $N_3=9$. We set $\mgm= 200$ GeV, which corresponds to
$F=(2.5\cdot 10^{10}\,{\rm GeV})^2$ and is of the same order as
$m_{3/2}\approx 150$ GeV by construction, and as $m_Z$. With this data, and with $\mu\simeq\sqrt{B_\mu}\simeq m_{3/2}$ suitably chosen, we can compute the 
low-scale particle spectrum at
two-loop precision using \verb!SOFTSUSY! \cite{Allanach:2001kg}. The lightest
Higgs mass is found to be $123.4$ GeV by \verb!SOFTSUSY! and $122.4$ GeV by
\verb!FeynHiggs! \cite{Frank:2006yh}. Taking into account a theoretical
uncertainty of at least $1-2$ GeV, this is marginally compatible with the region
currently favoured by ATLAS and CMS. Other features of the spectrum at the
electroweak scale include
\begin{itemize}
 \item large $\tan\beta\approx 50$, as a consequence of the fact that $B_\mu$ is
generated by gravity mediation but $m_{H_{u,d}}^2$ are dominated by gauge
mediation;
 \item a very heavy gluino, $M_3\approx 3.8$ TeV;
 \item very heavy squarks, the lightest of which is the $\tilde t_1$ at about
$2.5$ TeV, while the first-generation squarks are all heavier than $3$ TeV;
\item the remaining Higgs bosons $H^\pm$, $H^0$ and $A$ at intermediate masses,
at about $1.5$ TeV in our benchmark point;
\item a right-handed stau as the lightest scalar superparticle;
\item three light higgsinos whose mass scale is set by the gravity-mediated
$\mu$ parameter. As explained in some detail in \cite{Brummer:2011yd}, these are
two neutralinos and a chargino which are nearly degenerate in mass. Their most
natural mass range is about $150-300$ GeV, since the gravitino and the 
higgsinos should have comparable masses, with the gravitino being the LSP.
\end{itemize}

A similar but even slightly heavier spectrum is obtained with $N_1=N_2=28$, $N_3=11$. 
Spectra with $N_1<N_2$ can also be found. Interestingly,
upon decreasing $N_1$ the $\tilde\tau_1$ can become rather light, as a 
consequence of large $\tan\beta$ and the fact that the right-handed stau soft
mass is induced only by hypercharged messengers. For comparison, we have summarized the
masses of three benchmark spectra in Table \ref{tab:spectra}: the above one
with $(N_1,N_2,N_3)=(23,23,9)$; one with $(17,23,9)$ and a relatively light
$\tilde\tau_1$; and one with $(28,28,11)$.

\begin{table}
\begin{center}
\begin{tabular}{cccc}
particle & $(23,23,9)$ model & $(17,23,9)$ model & $(28,28,11)$ model\\ \hline
${h_0}$ &  $123$ & $123$ & $124$\\
${\chi^0_1}$ &  $205$ & $205$ & $164$\\
${\chi^\pm_1}$ &  $207$ & $206$ & $166$\\
${\chi^0_2}$ &  $208$ & $207$ & $167$\\
\hline
${\tilde\tau_1}$ & $1530$ & $550$ & $1890$\\ 
${H^0}$ & $1470$ & $1110$ & $2200$\\
${A}$ & $1480$ & $1120$ & $2200$\\
${H^\pm}$ & $1480$& $1120$ & $2200$\\ 
\hline
$\chi^0_3$ & $2500$& $1800$ & $2700$\\
$\chi^0_4$ & $3800$& $3800$ & $4100$\\
$\chi^\pm_2$ & $3800$& $3800$ & $4100$\\
\hline
${\tilde g}$ &  $3800$ & $3800$ & $4200$\\
${\tilde t_1}$ & $2500$ & $2300$ & $2700$\\ 
${\tilde u_1}$ & $3700$ & $3500$ & $4000$\\ 
${\tilde d_1}$ & $3400$ & $3400$ & $3700$
\end{tabular}
\end{center}
\caption{Some selected masses in GeV, computed with \texttt{SOFTSUSY}, 
for three models with messenger indices $(N_1,N_2,N_3)=(23,23,9)$, $(17,23,9)$, 
and $(28,28,11)$. The first has $\mgm=200$ GeV, $\mu=240$ GeV, and $\tan\beta=50$;
the second, $\mgm=200$ GeV, $\mu=250$ GeV, and $\tan\beta=52$; and the third,
$\mgm=180$ GeV, $\mu=180$ GeV, and $\tan\beta=44$.}\label{tab:spectra}
\end{table}

Admittedly, the choice of parameters described above is probably at the
boundary of what might be considered acceptable. Large messenger multiplicities
will lead to very large threshold corrections to gauge coupling unification, and
the theoretical uncertainties in the calculation of the spectrum are
considerable for multi-TeV soft masses. In addition, the Higgs mass is at the
lower end of the range indicated by experiment.

A spectrum like this comes close to a ``nightmare scenario'' for the LHC,
since all coloured particles are kinematically out of reach. The light higgsinos
will be produced in electroweak processes, but their decay signals are extremely
difficult to distinguish from Standard Model backgrounds
\cite{Baer:2011ec,Bobrovskyi:2011jj}. It may be possible to see a signal in
monojet searches, where a higgsino pair is produced in the Drell-Yan process and
shows up as missing $E_T$, while a gluon is radiated from the initial state.
The $\tilde\tau_1$ could also be within LHC reach if $N_1$ happens to
be sufficiently small, but will be similarly difficult to detect.
The situation will be much better at a linear collider, where all three higgsinos 
and possibly the $\tilde\tau_1$ can be produced and their masses and couplings can 
be accurately measured.   

\section{Fine-tuning}
One might object to our approach on the grounds that the coefficients in
Eq.~\eqref{eq:numericMZ} are disturbingly sensitive to variations in the Yukawa
couplings, the gauge couplings, and to the overall scale of SUSY breaking.
The worst sensitivity is with respect to the latter two: For example,
changing the value of $g^2(M)$ from $0.484$ to $0.5$ while keeping all other 
parameters fixed, Eq.~\eqref{eq:numericMZ} becomes
\be
\widehat m_Z^2=\,\left.\Bigl(2.50\,M_3^2-0.45\,M_2^2-0.01\,M_1^2+\ldots\Bigr)\right|_{M}
\ee
and after inserting the messenger numbers $(N_1,N_2,N_3)=(23,23,9)$, the
cancellation in Eq.~\eqref{eq:mZgauge1} is lost:
\be
\delGM= 26\,\mgm^2\,.
\ee
Similarly, keeping $g^2(M)=0.483$ but changing $\mS$ to $10$ TeV (which is of course
equivalent to an appropriate change of the background field $F$) gives
\be
\widehat m_Z^2 =\,\left.\Bigl(1.98\,M_3^2-0.45\,M_2^2-0.01\,M_1^2+\ldots\Bigr)\right|_{M}
\ee
and Eq.~\eqref{eq:mZgauge1} becomes
\be
\delGM= -31\,\mgm^2\,.
\ee
Is our focus point therefore merely relegating the fine-tuning to these
parameters? Or, to put it more drastically, would we still consider this
scenario natural in the hypothetical situation that the Standard Model gauge
couplings and the EWSB scale had not yet been measured?

While we cannot dispel these objections altogether, they might at least be 
mitigated by the following arguments. In the String Landscape
we expect all fundamental parameters to be discrete, and the fine-tuning problem
to disappear. The question is if we can realize a mechanism by which this
happens already in effective field theory. At present it seems impossible to
identify, within this framework, a physical principle which fixes the GUT-scale
gauge coupling to exactly its measured value, without introducing additional
free continuous parameters. The same is true for the SUSY breaking scale.
However, in realistic string models these two quantities are of course
correlated. In fact, in the heterotic models by which our models are inspired,
they are correlated in a rather precisely defined way: The gauge coupling is set
by the vacuum expectation value of the dilaton $S$,
\be
\frac{8\pi^2}{g^2}=S\,,
\ee
while the scale of dynamical SUSY breaking is set by the strong-coupling scale
$\Lambda$ of some non-abelian gauge group in the hidden sector, which is
ultimately related to the dilaton by
\be
\sqrt{F}\simeq\Lambda\simeq e^{-S/b_0}\,.
\ee
Decreasing $S$ will simultaneously increase $g(M)$ and $\mS$, and the resulting
effects on $\delGM$ can cancel (as should be clear from the above
example) for a suitable value of the discrete quantity $b_0$. Of course the
actual value of $b_0$ is model-dependent; nevertheless, this shows that it may
not be sensible to treat the gauge coupling and the SUSY breaking scale as
independent parameters.

On the other hand, changing the continuous parameters slightly can give new 
characteristic relations between the messenger indices which again lead to
a focus point at the electroweak scale. Empirically, we find that these always 
occur at a $N_3\,:\,N_2$ ratio of around $0.39$ (e.g.~for our models we have
$11/28\approx 0.393$ and $9/23\approx 0.391$). To determine which of them best 
describes our universe, it would however be necessary to measure the mass
ratios of the supersymmetric particles rather precisely --- a task which is clearly 
beyond the capabilities of the LHC for multi-TeV soft masses, 
but does not seem impossible in principle. 

\section{Conclusions}

In this note we have argued that the little hierarchy between the EWSB scale
and the MSSM soft masses may be due to a focus point exhibited by models of
high-scale gauge mediation. Since they are determined by discrete messenger
indices, the soft terms can be of the order of several TeV without large
fine-tuning. We have identified two favourable ratios of messenger indices
for $\SU 2$ and $\SU 3$ fundamentals, $(N_2,\,N_3)=(23,9)$ and $(28,11)$.
With these rather large multiplicities, the spectrum can be made
compatible with the recent experimental evidence for a $124-126$ GeV Higgs. 
Unfortunately, for this choice of parameters almost all superpartners are beyond 
the reach of the LHC. The only supersymmetric particles with weak-scale masses 
are the higgsinos and possibly the lightest stau, all of which would probably 
need a linear collider to be discovered, and the gravitino LSP.

\subsection*{Acknowledgements}

The authors thank G.~Ross for helpful discussions.

\end{document}